\documentclass[11pt, a4paper]{article}

\usepackage[utf8]{inputenc}
\usepackage{geometry}           
\usepackage{graphicx}           
\usepackage{amsmath}            
\usepackage{amsfonts}           
\usepackage{siunitx}            
\usepackage{xcolor}             
\usepackage{hyperref}           
\usepackage{authblk}            
\usepackage{tabularx}           
\usepackage{float}              
\usepackage{upgreek}            
\usepackage{caption}            
\usepackage{booktabs}

\hypersetup{
    colorlinks=true,
    linkcolor=blue,
    filecolor=magenta,
    urlcolor=cyan,
    pdftitle={Comparative Analysis of Thermal Models for Test Masses},
    pdfauthor={Vincenzo Pierro, et al.},
}

\newcommand{\parder}[2]{\frac{\partial #1}{\partial #2}}

\newcommand{\Omegam}{\Omega_m}
\newcommand{\Gammad}{\Gamma_d}

\title{\textbf{Comparative Analysis of Thermal Models for Test Masses in Next-Generation Gravitational Wave Interferometers}}

\author[1,2]{Vincenzo Pierro}
\author[2,3,*]{Vincenzo Fiumara}
\author[4]{Guerino Avallone}
\author[2,4]{Giovanni Carapella}
\author[2,5]{Francesco Chiadini}
\author[2,5]{Roberta De Simone}
\author[2,6]{Rosalba Fittipaldi}
\author[2]{Gerardo Iannone}
\author[7,8]{Alessandro Magalotti}
\author[7,8,*]{Enrico Silva}
\author[2,7]{Veronica Granata}

\affil[1]{Dipartimento di Ingegneria (DING), Università del Sannio, 82100 Benevento, Italy}
\affil[2]{INFN Sez. Napoli, Gruppo Collegato di Salerno, Italy}
\affil[3]{Dipartimento di Ingegneria, Università della Basilicata, 85100 Potenza, Italy}
\affil[4]{Dipartimento di Fisica “E.R. Caianiello”, Università di Salerno, 84084 Fisciano SA, Italy}
\affil[5]{Dipartimento di Ingegneria Industriale (DIIN), Università di Salerno, 84084 Fisciano SA, Italy}
\affil[6]{Istituto Spin-CNR, 84084 Fisciano SA, Italy}
\affil[7]{Dipartimento di Ingegneria Industriale, Elettronica e Meccanica (DIIEM), Università di Roma Tre, 00146 Roma, Italy}
\affil[8]{INFN Sez. Roma Tre, 00146 Roma, Italy}
\affil[*]{\textit{Corresponding authors:} vincenzo.fiumara@unibas.it (V.F.); enrico.silva@uniroma3.it (E.S.)}

\date{October 9, 2025} 

\begin{document}

\maketitle

\begin{abstract}
Accurate thermal modeling of Terminal Test Masses (TTMs) is crucial for optimizing the sensitivity of gravitational wave interferometers like Virgo. In fact, in such gravitational wave detectors even minimal laser power absorption can induce performance-limiting thermal effects. This paper presents a detailed investigation into the steady-state thermal behavior of TTMs. In particular, future scenarios of increased intracavity laser beam power and optical coating absorption are considered. We develop and compare two numerical models: a comprehensive model incorporating volumetric heat absorption in both the multilayer coating and the bulk substrate, and a simplified reduced model where the coating's thermal impact is represented as an effective surface boundary condition on the substrate. Our simulations were focused on a ternary coating design, which is a candidate for use in next-generation detectors. Results reveal that higher coating absorption localizes peak temperatures near the coating--vacuum interface. Importantly, the comparative analysis demonstrates that temperature predictions from the reduced model differ from the detailed model by only milli-Kelvins, a discrepancy often within the experimental uncertainties of the system's thermo-physical parameters. This finding suggests that computationally efficient reduced models can provide sufficiently accurate results for thermal management and first-order distortion analyses. Moreover, the critical role of accurately characterizing the total power absorbed by the coating is emphasized.
\end{abstract}

\vspace{1em}

\section{Introduction}
\label{sec:introduction}

The quest for detecting gravitational waves, ripples in the fabric of spacetime predicted by Einstein's theory of General Relativity, has spurred the development of extraordinarily sensitive interferometric detectors such as LIGO, Virgo, and KAGRA~\cite{Virgosite,LIGOsite, KAGRA}. These instruments operate by measuring minuscule changes in the relative path lengths of laser beams traveling along kilometer-scale arms, induced by passing gravitational waves. The core optical components of these interferometers, particularly the Terminal Test Masses (TTMs) that form the end mirrors of the Fabry--Pérot cavities, are subjected to extremely high circulating laser powers, up to the megawatt range in future upgraded designs~\cite{Virgo_Advanced, LIGO_Advanced, LVK}.

Even minute absorption of this high optical power within the TTMs—both in their bulk substrate material and, more critically, in their highly reflective optical coatings—can lead to significant thermal effects. The importance of simulating and understanding these thermal phenomena was recognized in many studies, in particular, for thermal compensation, let us cite the foundational studies by Hello and Vinet \cite{HV1990ThermalAberrations, HelloVinet1990JPF,Hello1993, Hello1993JPF} (see~\cite{Rocchi2012} for an update).

The thermal effects manifest mainly as (i) thermal noise, including coating Brownian noise, thermo-elastic noise, and thermo-refractive noise~\cite{HarryBook}, which can mask faint gravitational wave signals~\cite{SaulsonBook}; (ii) thermal lensing, where temperature gradients alter the refractive index of the optical materials; (iii) physical deformation of the mirror surfaces. The combination of these effects degrades the interferometer's optical performance and stability~\cite{VirgoTCS}, prompting extensive research into methods for their compensation~\cite{Hello2001EPJD} and the development of dedicated thermal compensation systems, as exemplified by early designs for Virgo~\cite{Accadia1}. Consequently, accurate modeling and understanding of the temperature distribution within TTMs under operational conditions are paramount for predicting, mitigating, and compensating for these deleterious thermal effects, ultimately pushing the sensitivity limits of gravitational wave detection.

Optical coatings, typically composed of multilayer dielectric stacks, are directly exposed to the incident laser light and thus become a significant source of both optical absorption and thermal noise. The physical thickness of these coatings is typically very small (on the order of micrometers) compared to the dimensions of the bulk substrate (tens of centimeters). However, their complex layered structure and the thermo-mechanical properties of the thin-film materials---which can differ significantly from their bulk counterparts---pose considerable challenges for accurate thermal modeling~\cite{ThermCompl}.

This paper presents a detailed investigation into the thermal behavior of TTMs, where the highly reflective coating is realized by using a stack of three different materials (ternary coating)~\cite{VPierro}. These types of coatings are of interest because they have been shown to outperform, in terms of Brownian noise reduction, the coatings currently used in gravitational wave detectors, which are made of two materials (binary coating)~\cite{Tantala}. The goal of the paper is to present the first thermal analysis, to the best of our knowledge, of MMTs involving ternary optical coating.

The paper is focused on two key aspects. Firstly, we develop and implement a comprehensive thermal model that explicitly accounts for volumetric heat absorption within both ternary multilayer optical coatings and bulk substrates, coupled with radiative heat exchange with the environment. This model is employed to analyze the steady-state temperature distribution, with a particular emphasis on scenarios involving higher coating absorption levels than is traditional. Such conditions arise from two key factors in next-generation interferometers: the use of novel coating materials and an increase in intracavity laser beam power~\cite{CExplorer}.

Secondly, recognizing the computational complexity associated with resolving the thermal field within the microscopically thin yet structurally intricate coating, we explore the viability of a simplified, reduced thermal model. This reduced model aims to represent the entire thermal effect of the coating (including its internal heat generation and surface heat exchange) as an effective boundary condition applied directly to the substrate at the coating--substrate interface. We conduct a rigorous comparative analysis between the detailed volumetric model and this reduced model, evaluating the discrepancies in their temperature predictions both axially and radially. The goal is to assess whether such a simplification, which offers significant computational advantages, can provide sufficiently accurate results for practical thermal management and first-order distortion analyses, especially in light of inherent uncertainties in experimentally determined thermo-physical~parameters.

The structure of this paper is as follows: Section~\ref{sec:thermal_model} details the formulation of the comprehensive thermal model for the TTM, including the geometry, material properties, heat source distributions, and boundary conditions. Section~\ref{sec:heat_generation_substrate} and Section~\ref{sec:heat_generation_coating} describe the models for heat generation within the substrate and the homogenized coating, respectively. Section~\ref{sec:simulation_results} presents the simulation results obtained using the detailed model for a specific ternary coating design, illustrating the temperature profiles and key thermal characteristics. Section~\ref{sec:reduced_model_comparison} introduces the reduced thermal model and presents a comparative analysis of its predictions against the detailed model. Finally, Section~\ref{sec:conclusions} summarizes the main findings of this work and discusses their implications for the thermal design and modeling of future gravitational wave interferometers.

\section{A Thermal Model of a Gravitational Wave Interferometer Terminal Test Mass}
\label{sec:thermal_model}

In this section we give a description of the thermal model for a Terminal Test Mass (TTM) within an advanced gravitational wave interferometer cavity.

\begin{figure}[H]
	\centering
	\includegraphics[width=0.8\textwidth]{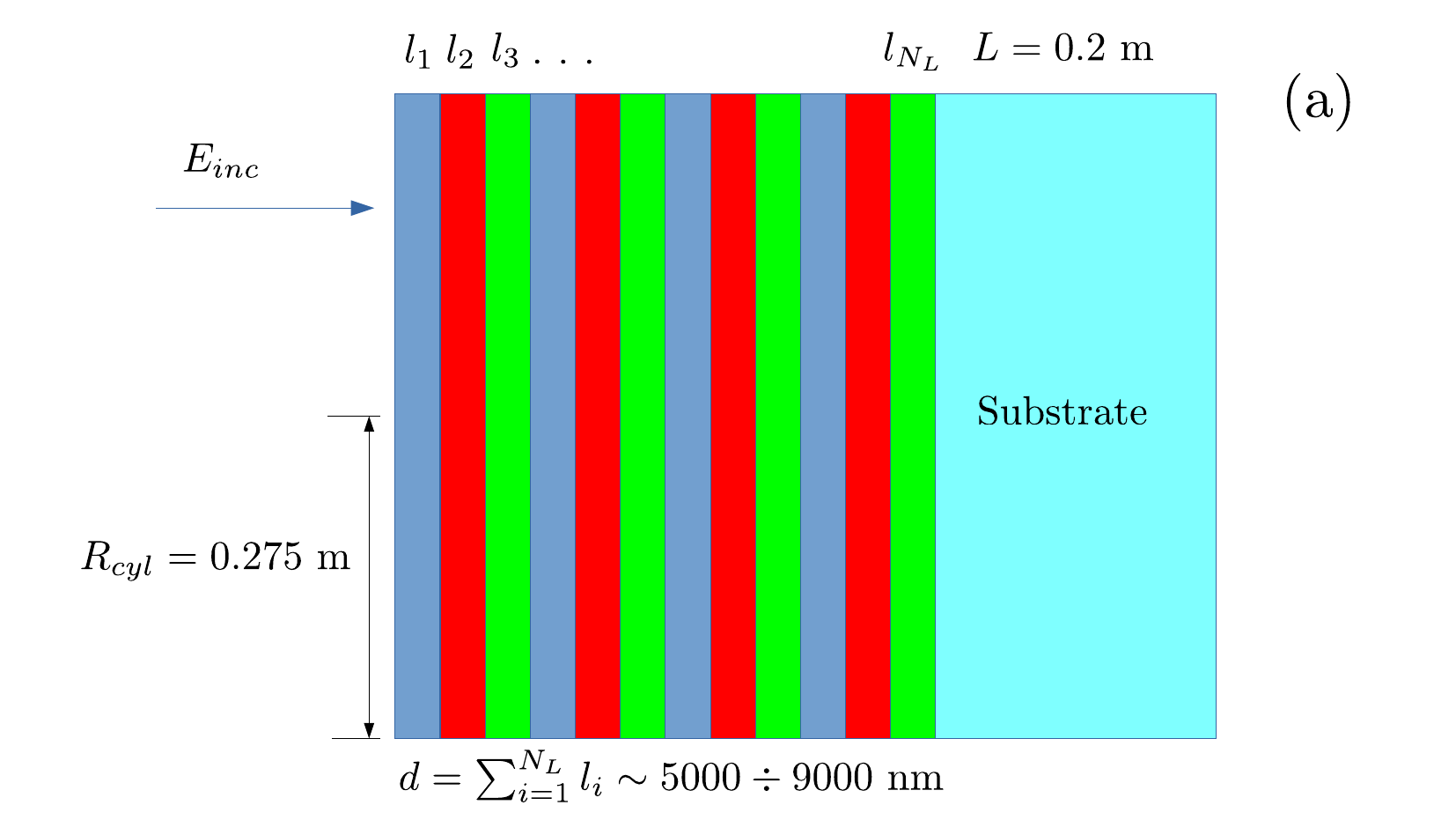}\\
	\includegraphics[width=0.8\textwidth]{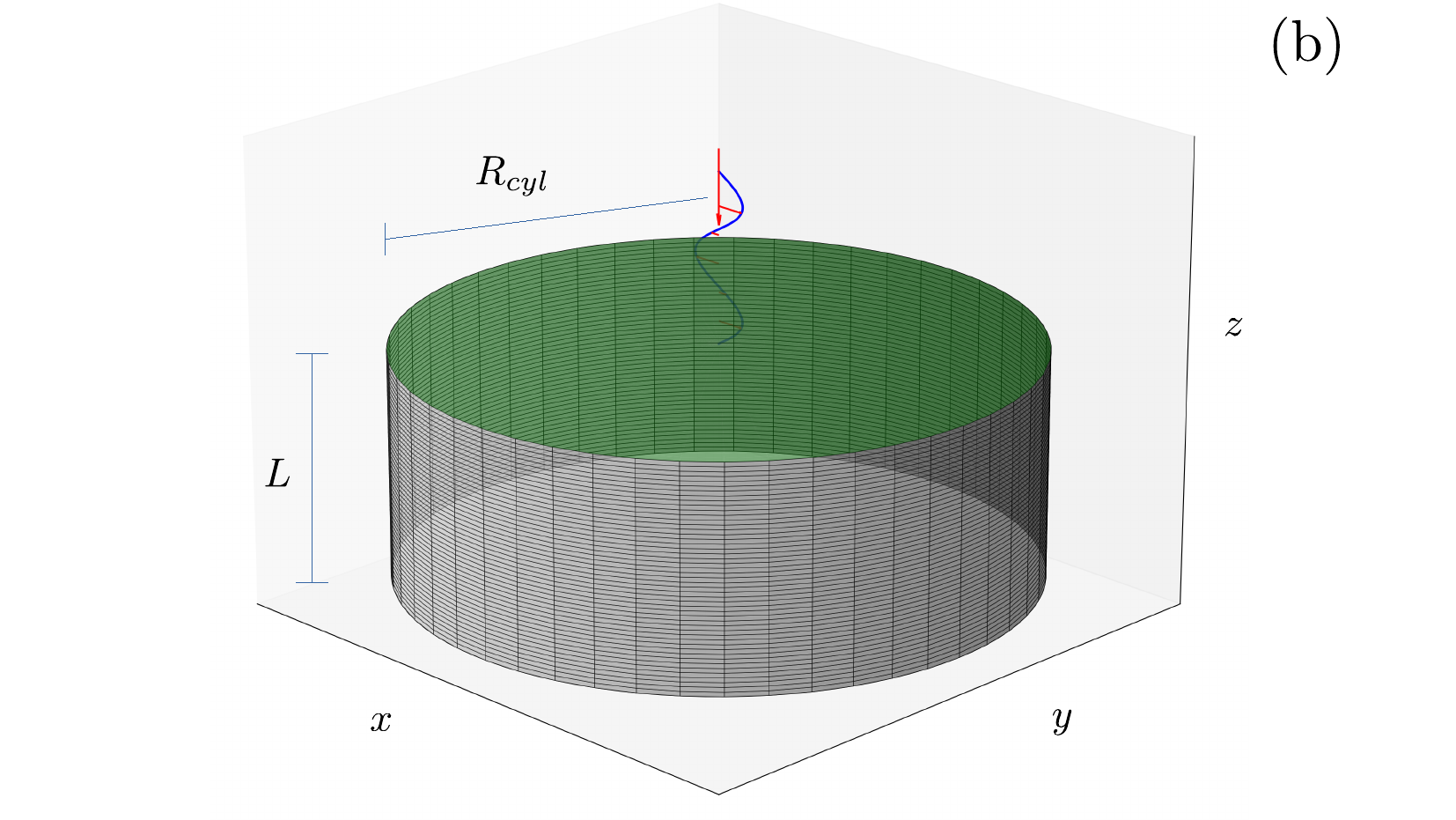}
	\caption{Panel (\textbf{a}) shows a cross-sectional view of the test mass, specifically focusing on the top surface where the incident electromagnetic field, $E_{inc}$, strikes. Different colors corespond to different materials. Panel (\textbf{b}) provides a 3D perspective of the test mass, which is a cylinder. Its dimensions are indicated: radius $R_{cyl}$ and height $L+d$, where $L$ is the substrate thickness. The top face of the cylinder is colored green, representing the surface where the coating (detailed in panel (\textbf{a})) is deposited. The incident electromagnetic field (i.e., Gaussian beam) is propagating towards the coated top surface.}
	\label{fig:panel} 
\end{figure}

A schematic of a standard TTM is shown in Figure~\ref{fig:panel}. Panel (a) shows a cross-sectional view of the TTM, which consists of a cylindrical substrate coated with a multilayer reflector. This view is not to scale, and the layers of the coating are actually very thin compared to the substrate dimensions. The coating is a layered structure consisting of a sequence of different materials with individual thicknesses denoted by $l_1, l_2, l_3, \dots$ up to $l_{N_L}$ ($N_L$ is the total number of layers). The total thickness of the coating, $d = \sum_{i=1}^{N_L} l_i$, is generally between $5000~\mathrm{nm}$ and $9000~\mathrm{nm}$. Panel (b) provides a 3D perspective of the test mass, which is a cylinder with a radius $R_{cyl} = 0.275~\mathrm{m}$ and total height $L+d$, where $L = 0.2~\mathrm{m}$ is the thickness of the substrate. The coating is deposited on the top face of the cylinder (colored green in the figure). There, a time-harmonic electromagnetic Gaussian beam with a wavelength of $\lambda=1064$\,nm is normally incident, which, for the optical simulations, we approximate as a plane wave described by the harmonic phasor $E_{inc}$.

This section presents a thermal model for a TTM under the simplifying assumption of an axisymmetric cylindrical geometry. The model encompasses a multi-component structure, incorporating distinct heat absorption profiles within both the optical coating and the bulk substrate, and considers radiative heat exchange with the surrounding environment.

A TTM is not a simple monolithic structure; it comprises two principal regions:

\begin{itemize}
    \item \textbf{The Bulk Substrate:} This is the main body of the test mass, extending axially from $z=d$ to $z=L+d$ (see Figure~\ref{fig:panel}b). It is modeled as a homogeneous material with thermal conductivity $k_m$. The region occupied by the substrate is denoted by $\Omega_m$.
    \item \textbf{The Optical Coating:} Applied to one face of the substrate (the region $0 < z < d$, depicted in Figure~\ref{fig:panel}a), this is a highly reflective multilayer structure with average thermal conductivity $k_c$. This coating is the primary point of interaction with the high-power intracavity laser beam. Despite its physical thinness ($d << L$), its thermal behavior is crucial. The region occupied by the coating is denoted by $\Omega_c$.
\end{itemize}

A key aspect of our model is the detailed characterization of heat deposition due to the incident radiation. We consider two distinct, spatially varying heat sources, both sharing a Gaussian radial profile, which is due to the incident Gaussian laser beam:

\begin{itemize}
    \item \textbf{Coating Absorption ($q_c$):} The dominant heat deposition occurs within the coating due to optical absorption. This is meticulously modeled as a volumetric heat source: $q_c(r,z) = A_0 \exp(-2 r^2/w^2) \exp(-\alpha_0 z)$. Here, $A_0$ represents the peak power absorption density, $w$ is the Gaussian beam radius (waist) at the vacuum--coating interface, and $\alpha_0$ characterizes the exponential decay of absorption with depth $z$ into the coating.
    \item \textbf{Substrate Absorption ($q_m$):} {A secondary volumetric heat absorption within the bulk substrate is also included. This is modeled as $q_m(r,z) = A_1 \exp(-2 r^2/w^2) \exp(-\alpha_1 (z-d))$} for $z>d$, where $A_1$ and $\alpha_1$ are the corresponding peak absorption and decay constants for the substrate material. The shared Gaussian beam waist $w$ signifies that this absorption is also driven by the main laser beam, due to residual transmission through the coating.
\end{itemize}

The radial Gaussian profile is determined by the radiation incident on the $z=0$ face, i.e., a Gaussian beam with a waist equal to $w$. Thermal exchange between the test mass and the environment is governed by radiative heat transfer. This radiative condition is simplified to an effective convective condition, given the expected low temperature gradient between the surface and the surroundings. 

Our model incorporates the following boundary conditions:
\begin{itemize}
    \item The outer face of the coating (at $z=0$) exchanges heat with its surroundings; the linearized heat radiative transfer coefficient is $h_0=4 \epsilon_s \sigma_{SB} T_\infty^3$, where $T_{\infty}$ is the ambient (room) temperature, $\epsilon_s$ the emissivity, and $\sigma_{SB}$ the Stefan--Boltzman constant. The governing equation is
    \begin{equation*}
        -k_c \parder{T_c}{z}\Big|_{z=0} = - h_0 (T_c(r,0) - T_{\infty}),
    \end{equation*}
    where $T_c$ is the temperature field in the coating.
    \item Similarly, the opposing face of the substrate (at $z=L+d$) and the cylindrical lateral surface (at $r=R_{cyl}$) are also subject to linearized radiative boundary conditions, with the same heat transfer coefficients and ambient temperatures ($h_0, T_{\infty}$). For instance, at $z=L+d$,
    \begin{equation*}
        -k_m \parder{T_m}{z}\Big|_{z=L+d} = h_0 (T_m(r,L+d) - T_{\infty}).
    \end{equation*}
    Here, $T_m$ is the temperature field in the substrate.
    \item Axial symmetry is, of course, enforced at $r=0$: $\parder{T}{r}\Big|_{r=0} = 0$.
\end{itemize}

For bulk silica used as optical substrates, the typical emissivity of $\epsilon_s \sim 0.8$ implies a heat transfer coefficient of $h_0 \sim 4.8$~W$\cdot$ m$^{-2}$$\cdot$ K$^{-1}$) at room temperature. We assume this value also applies to the outer face (at $z=0$) of the silica thin-film coating.

The main goal of this thermal model is to predict the steady-state temperature distribution, $T(r,z)$, throughout the test mass under these operational heat loads and cooling conditions. This detailed temperature map is the foundation for subsequent analyses, including calculating thermo-elastic deformation, estimating thermally induced optical path distortions, and predicting various forms of thermal noise that can limit the interferometer's~sensitivity.

Due to the multi-material nature, the thinness of the coating, and the complex interplay of heat generation and dissipation, accurately capturing the thermal gradients, especially at the coating--substrate interface ($\Gammad$ at $z=d$), necessitates careful numerical treatment.

\section{Heat Generation Inside the Substrate}
\label{sec:heat_generation_substrate}

Following its interaction with the multilayer coating, a portion of the incident Gaussian laser beam is transmitted into the substrate. We model this transmitted radiation as a traveling wave. The intensity profile of this transmitted beam retains its Gaussian spatial distribution due to the nature of the incident Gaussian beam. Therefore, the power density at a radial position $r$ and depth $z$ within the substrate is proportional to
\begin{equation}
\exp\left(-\frac{2 r^2}{w^2}\right) \exp(-\alpha_1 (z-d)).
\label{eq:substrate_intensity}
\end{equation}

Equation (\ref{eq:substrate_intensity}) is based on the assumption of a traveling wave model that solely considers the forward-propagating wave, with no reflections.

The heat generated per unit volume in the substrate, $q_s(r,z)$, is then directly proportional to the optical absorbed power density, and we have the following model equation,
\begin{equation}
q_s(r,z) = A_1 \exp\left(-\frac{2 r^2}{w^2}\right) \exp(-\alpha_1 (z-d))
\label{eq:substrate_heat_generation}
\end{equation}
where
\begin{itemize}
    \item $\exp\left(-\frac{2 r^2}{w^2}\right)$ describes the Gaussian spatial profile of the transmitted laser beam, where $w$ is the beam radius (assuming it remains approximately constant through the propagation in the substrate, see the Appendix for a discussion).
    \item $\alpha_1$ is the linear absorption coefficient of the substrate material at the laser wavelength. For a substrate made of Suprasil silica, $\alpha_1$ is determined from the imaginary part (i.e., extinction $k$) of its complex refractive index ($n = n_r - ik$). The relationship is $   \alpha_1 = \frac{4\pi k}{\lambda}$, where $\lambda$ is the vacuum wavelength of the laser radiation. 
    \item $\exp(-\alpha_1 (z-d))$ represents the exponential decay of the beam's power density as it propagates and is absorbed within the substrate, according to the Beer--Lambert law.
    \item The constant $A_1$ is related to the power entering the substrate. Let $P_{inc}$ be the power incident on the coating/cavity interface and $\tau_c$ be the transmittance of the entire coating stack. The total power transmitted through the coating and potentially available for absorption in the substrate is $P_s = \tau_c P_{inc}$. As a crude approximation, only the absorbed power due to the direct traveling wave propagating from the coating/substrate interface to the bottom of the substrate is considered. In order to calculate $A_1$, this approximation is modeled by defining an effective incident power $P_{eff\_inc}$:
	\begin{equation}
    P_{eff\_inc} = \tau_c P_{inc}\{1 - \exp[-\alpha_1 (L - d)]\}.
    \label{eq:P_eff_inc}
	\end{equation}
	The coefficient $A_1$ is then computed by imposing the following equation:
	\begin{equation}
    \int_{\Omegam} q_s(r,z) \, dV = P_{eff\_inc}
	\end{equation}
\end{itemize}

A key assumption in this model is the presence of an ideal anti-reflective (AR) matching layer at the bottom surface of the substrate. This AR coating is presumed to perfectly eliminate reflections at the substrate--vacuum interface at the exit face of the substrate. This simplifies the model by ensuring that all light reaching the bottom of the substrate exits without being reflected back into the substrate, thus preventing the formation of standing waves or multiple-pass absorption effects within the substrate itself. This allows for a straightforward traveling wave absorption model. The emissivity of the AR coating on the substrate's back face was assumed to be equal to that of the bulk silica, an approximation justified by the similar emissivity of its constituent dielectric materials (typically, MgF$_2$ and Ta$_2$O$_5$) in the relevant thermal infrared range.

\section{Heat Generation Inside the Coating}
\label{sec:heat_generation_coating} 

\begin{figure}[H]
	\centering
	\includegraphics[width=0.8\textwidth]{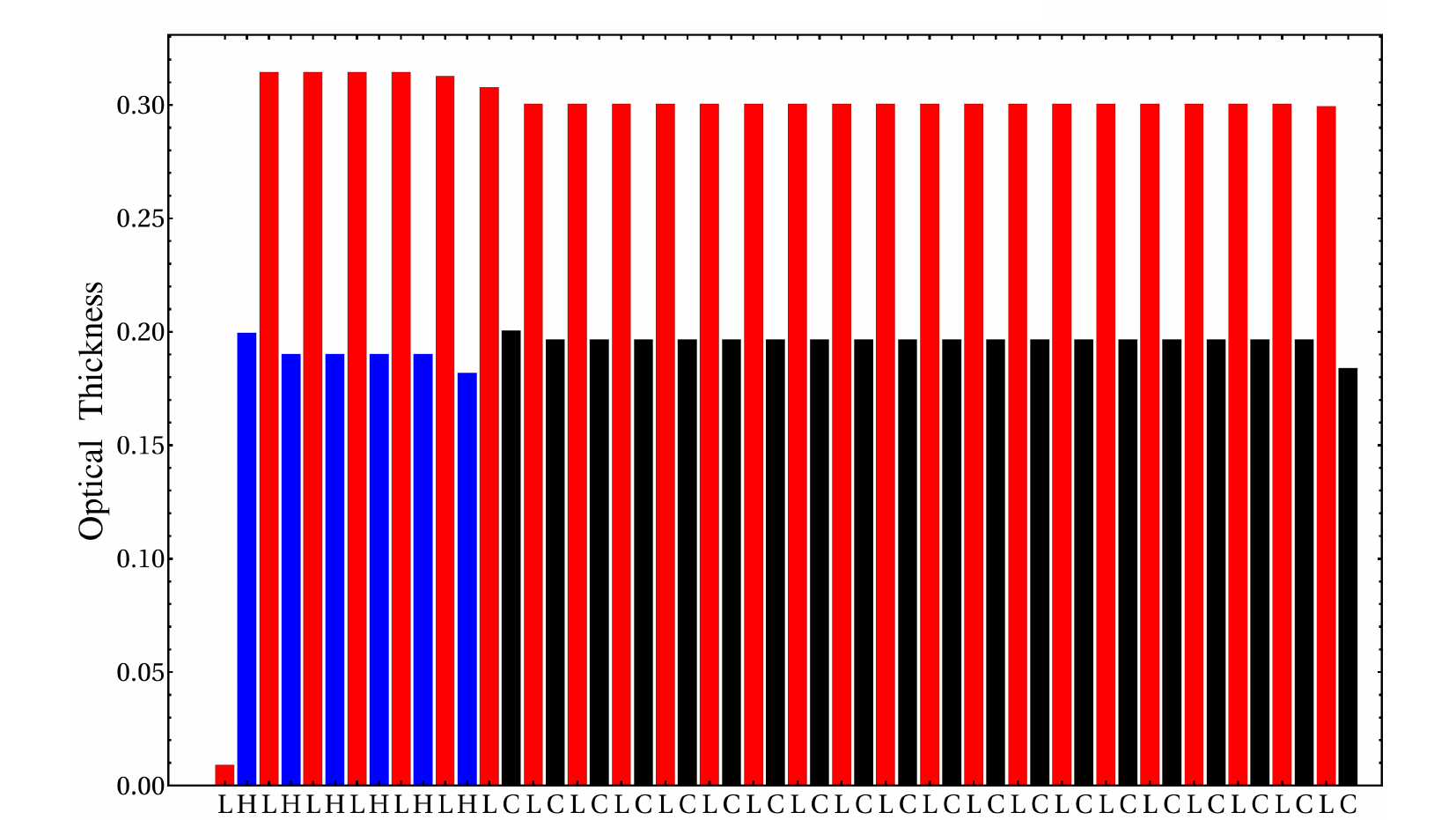}
	\caption{An example of a DSD coating structure made of Silica (denoted by ``L''), Ti::SiO$_2$, (denoted by ``H''), and Ti::GeO$_2$ (denoted by ``C''). The sequence of materials and the optical length of each layer are~displayed.}
	\label{fig:coating}
\end{figure}

A multilayer optical coating is typically composed of a sequence of thin-film layers, each characterized by a specific material and thickness (see the general structure in Figure~\ref{fig:panel}). As illustrated in Figure~\ref{fig:panel}a, these layers are stacked one upon another to achieve desired optical properties. The total thickness of such a coating can range from approximately $5000$ to $9000$ nanometers, for the specific application of highly reflective mirrors for gravitational detectors. Here we consider the ternary coating shown in Figure~\ref{fig:coating} that is a Double Stack of Doublets (DSD) design made of Silica/Ti::SiO$_2$/Ti::GeO$_2$ (see Table~\ref{tab:tabcond} for relevant parameters). This particular structure consists of an initial sequence of Silica/Ti::SiO$_2$ doublets, followed by a second sequence composed of Silica/Ti::GeO$_2$ doublets~\cite{VPierro}.

\begin{table}[H]
	\centering
	\caption{Estimated thermal conductivity ranges at room temperature and refractive indexes of the materials deposited in the coating and the substrate. The thermal conductivity values of the layers vary greatly due to the different deposition systems and conditions used to generate them, which can affect this parameter.}
	\label{tab:tabcond}
	\begin{tabularx}{\textwidth}{@{} >{\raggedright\arraybackslash}X >{\raggedright\arraybackslash}X >{\raggedright\arraybackslash}X >{\raggedright\arraybackslash}X @{}}
	\toprule
	\textbf{Material} & \textbf{Type} & \textbf{Thermal Conductivity (W} \boldmath{$\cdot$} \textbf{m}\boldmath{$^{-1}$ $\cdot$} \textbf{K}\boldmath{$^{-1}$}\textbf{)} &  \textbf{Refractive Index (@} \boldmath{$\lambda=1064$} \textbf{nm)} \\
	\midrule
	Bulk silica  & Amorphous &  1.3--1.5 &  $1.45 - \mathrm{i}\, 3\times 10^{-8}$\\
	SiO$_2$ & Amorphous & 1.0--1.35  & $1.45 - \mathrm{i}\, 3\times 10^{-8}$\\
	Ti::SiO\textsubscript{2} 62\%  & Amorphous sputtered film & 0.8--1.0 & $1.92 - \mathrm{i}\,  10^{-7}$ \\
	Ti::GeO\textsubscript{2} 44\%  & Amorphous sputtered film & 0.7--1.0 & $1.89 - \mathrm{i}\, 1.9\times 10^{-7}$\\
	\bottomrule
	\end{tabularx}
\end{table}

\begin{figure}[H]
	\centering
	\includegraphics[width=0.8\textwidth]{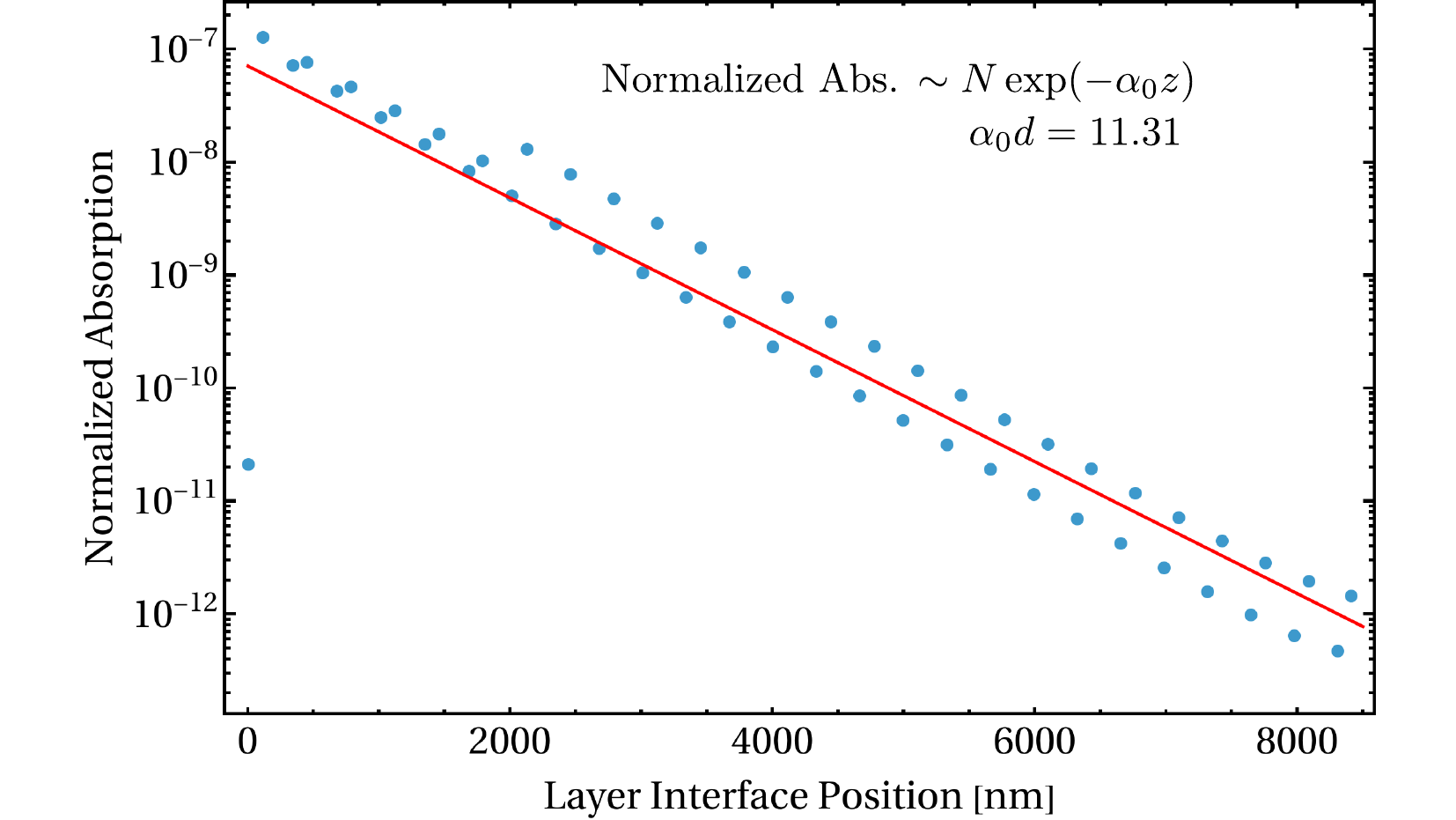}
	\caption{The normalized power absorption per layer (dots) and its continuous exponential fit (line). The model fit is exponential $\sim \exp(-\alpha_0 z)$, and the normalized value of the $\alpha_0$ parameter is displayed in the legend.}
	\label{fig:power}
\end{figure}

Figure~\ref{fig:power} presents the power absorbed within each layer of the coating, normalized to the total incident power computed by solving the coating--substrate optical boundary value problem, thus yielding the profile of the standing wave in each layer of the coating. 

For the purpose of thermal modeling, we homogenize the coating, i.e., we replace the discrete power generation profile (where power is deposited layer by layer) with a continuous, exponentially decaying function. This exponential decay, $\exp(-\alpha_0 z)$, is derived from a linear fit to the discrete power absorption data when plotted on a semi-logarithmic scale (logarithm of power vs. layer depth/position $z$), as indicated by the (red) line in Figure~\ref{fig:power}.

Consequently, the volumetric heat generation rate, $q_c(r,z)$, within the homogenized coating is modeled as
\begin{equation}
q_c(r,z) = A_0 \exp\left(-\frac{2 r^2}{w^2}\right) \exp(-\alpha_0 z)
\label{eq:heat_generation}
\end{equation}
where
\begin{itemize}
    \item $\alpha_0$ is the effective absorption coefficient for the homogenized coating, estimated from the slope of the linear fit to the power absorption data (as shown in Figure~\ref{fig:power}).
    \item $A_0$ is an integration constant. It is calculated by equating the total power absorbed in this continuous approximation to the total power absorbed in the discrete case (i.e., the sum of the normalized powers for each layer from Figure~\ref{fig:power} multiplied by the incident~power).
\end{itemize}

\section{Simulation Results: DSD Ternary SiO$_2$/Ti::SiO$_2$/Ti::GeO$_2$}
\label{sec:simulation_results} 

In this section, we report on the results for a SiO\textsubscript{2}/Ti::SiO\textsubscript{2}/Ti::GeO\textsubscript{2} ternary coating. The computational methods, detailed in~\cite{VP2,VP3}, specifically include the calculation of the electromagnetic field distribution within the coating using the Abel\'{e}s matrix method~\cite{Abele}. The numerical solutions of the heating problem were obtained using the finite element method, implemented within the open source software FEniCS~\cite{fenics_paper}.

We refer to this coating only as an example. The results obtained can be generalized to any type of coating used as a high-reflectivity mirror in interferometers for graviational wave detection. The coating (a multilayered structure) is $d=8415$ nm thick and is deposited on one of the bases of a silica bulk cylinder with a radius of $R_{cyl} = 0.275$ m and a height of $L = 0.2$ m (see Figure~\ref{fig:panel}b). The homogenization of the heat source in the coating, as described above, yields $\alpha_0 d= 11.31$; furthermore, we assume that $k_c = 0.8$ W$\cdot$ m$^{-1}$ $\cdot$ K$^{-1}$. 

The thermal conductivity values of materials in the form of thin layers exhibit significant variability, depending on the deposition system and the conditions employed in their fabrication process. The value used in our simulations, i.e., $k_c=0.8$ W$\cdot$ m$^{-1}$ $\cdot$ K$^{-1}$, is an average value weighted by the volume percentages of the various layers of materials.

\begin{figure}[H]
	\centering
	\includegraphics[width=0.8\textwidth]{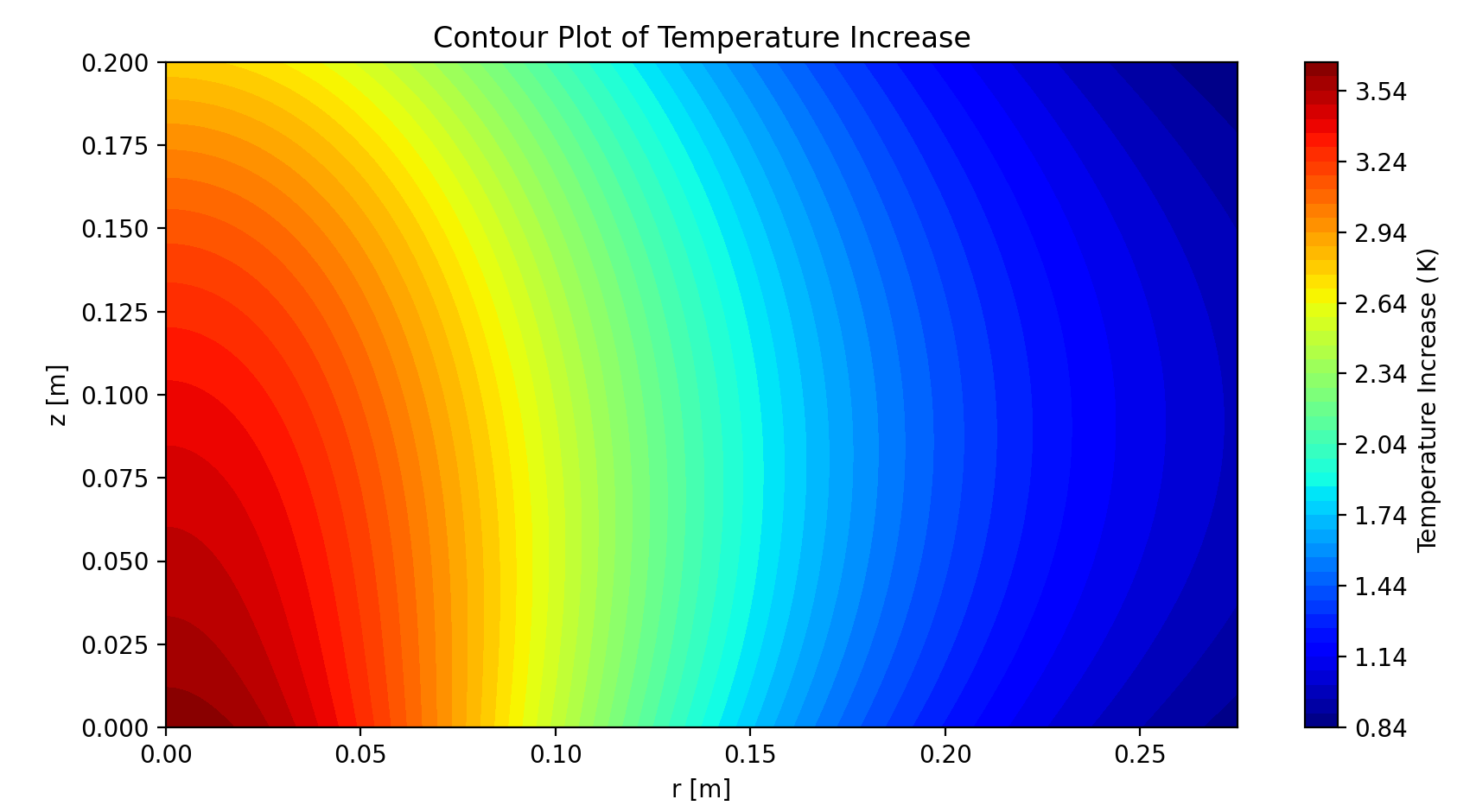}
	\caption{Increase in temperature, with respect to room temperature $T_\infty$, on the radial section of the~TTM.}
	\label{fig:contour}
\end{figure}

\begin{figure}[H]
	\centering
	\includegraphics[width=0.8\textwidth]{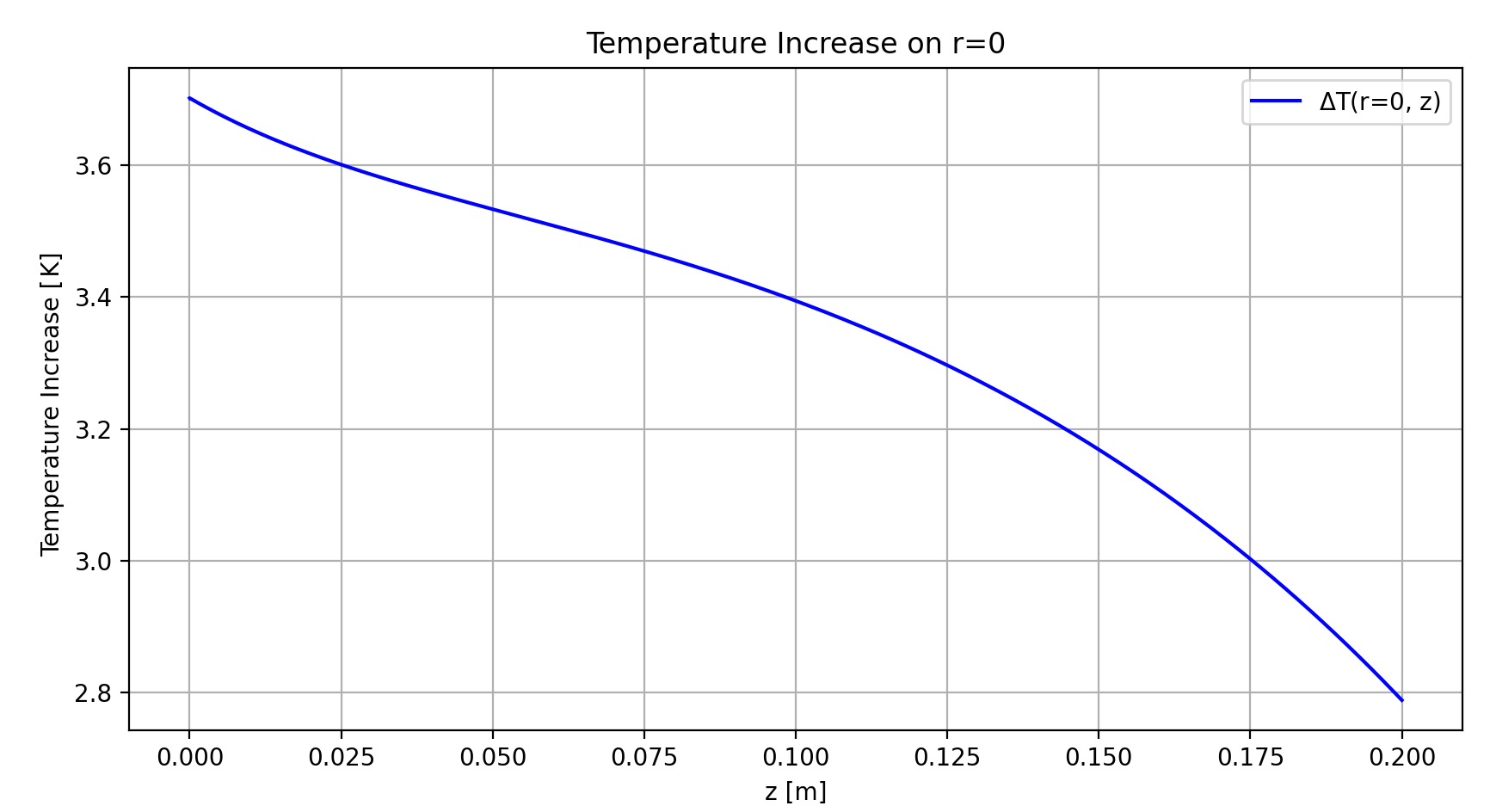}
	\caption{Increase in temperature ($\Delta T$), with respect to room temperature $T_\infty$, on the $r=0$ axis ($z \in [0,L]$).}
	\label{fig:increase1}
\end{figure}

\begin{figure}[H]
	\centering
	\includegraphics[width=0.9\textwidth]{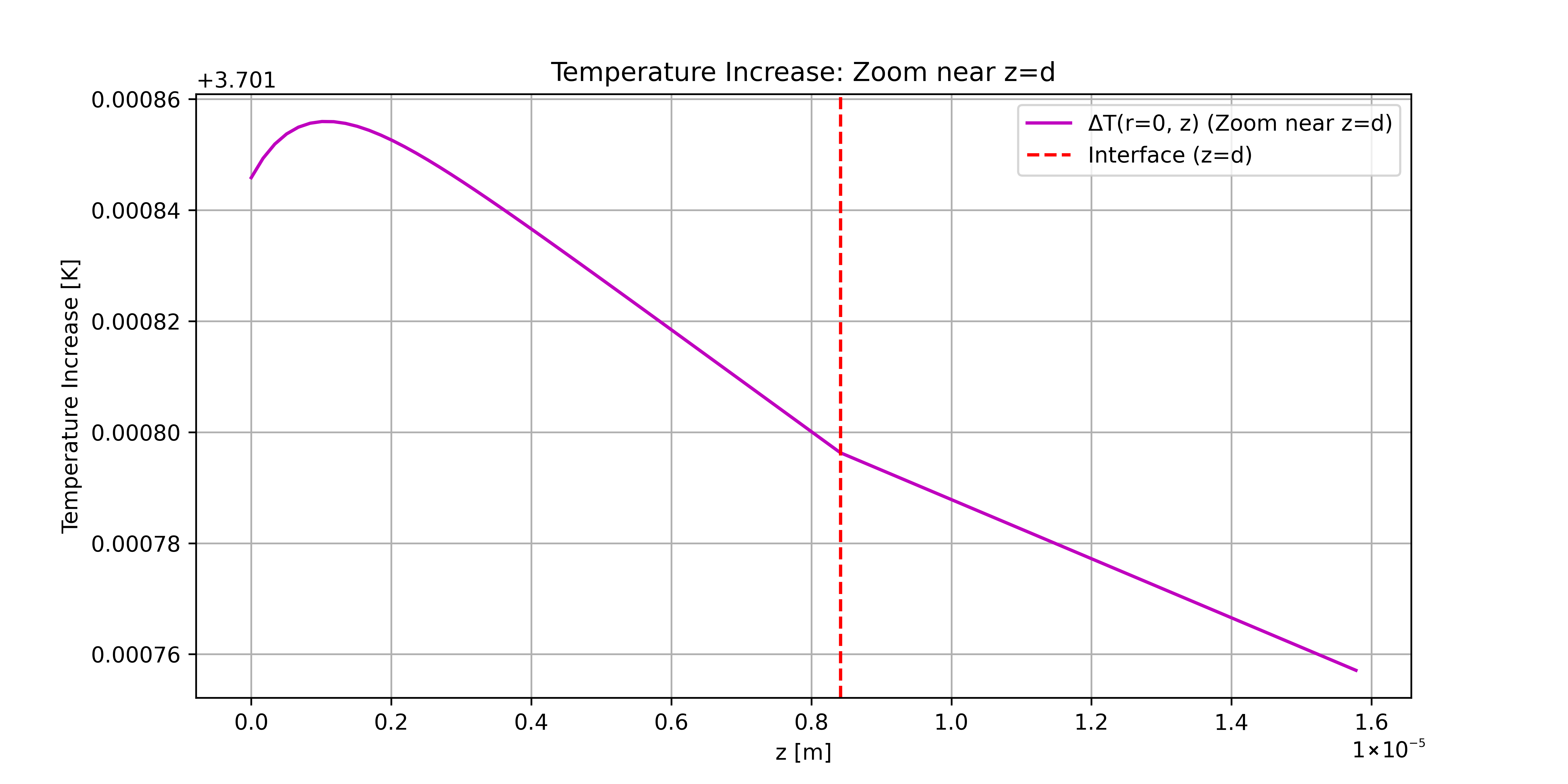}
	\caption{Increase in temperature ($\Delta T$), with respect to room temperature $T_\infty$, on the $r=0$ axis (close-up near the $z=d$ interface).}
	\label{fig:closeup}
\end{figure}

\begin{figure}[H]
	\centering
	\includegraphics[width=0.8\textwidth]{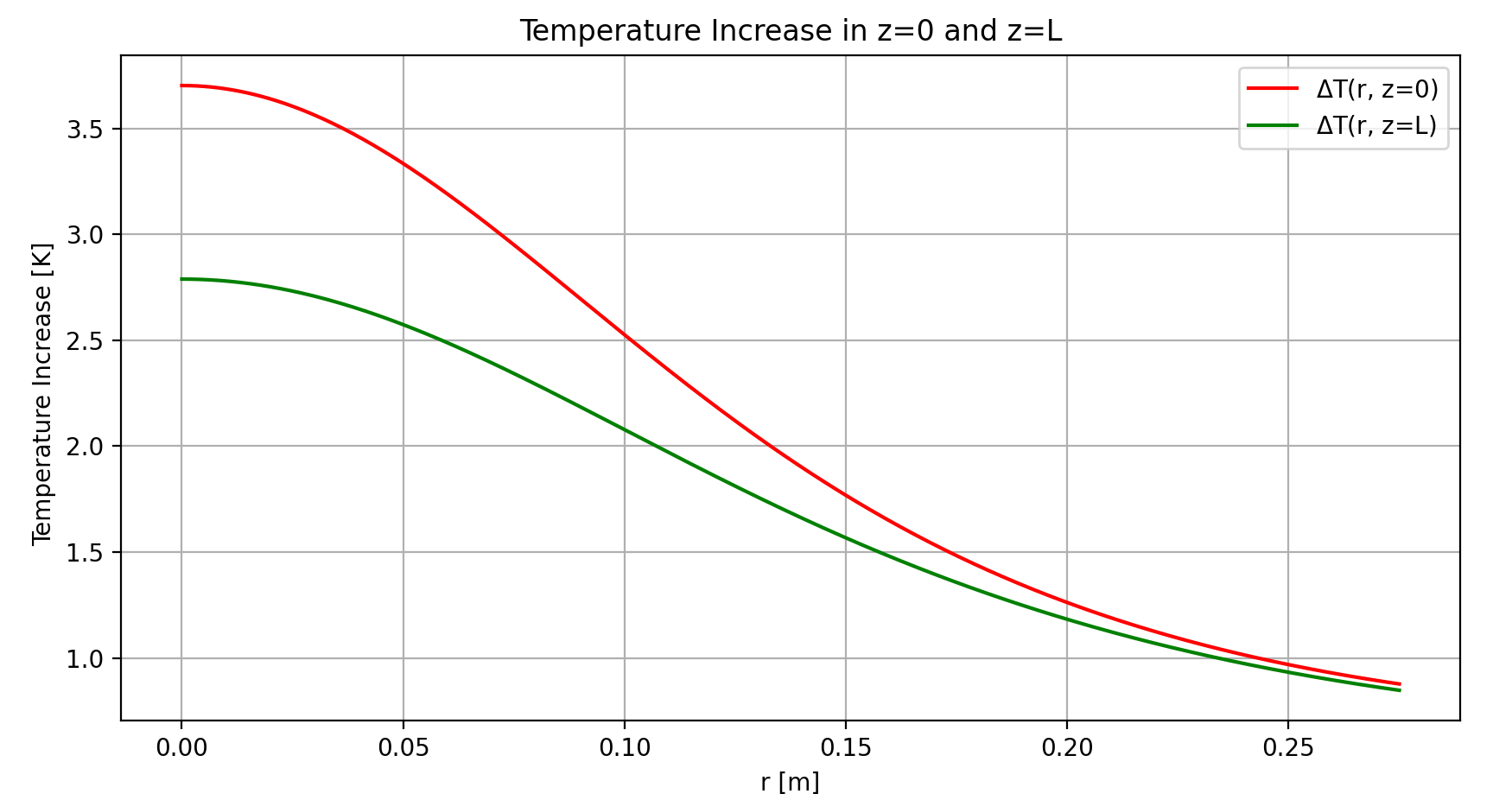}
	\caption{Increase in temperature ($\Delta T$), with respect to room temperature $T_\infty$, on the upper (red) $z=0$ and lower (green) $z=L$ cylinder faces.}
	\label{fig:increase2}
\end{figure}

On the other hand, the bulk silica in the substrate exhibits an extinction coefficient of $\kappa=3 \times 10^{-8}$, giving $\alpha_1 = 0.354\,$ m$^{-1}$\, and a thermal conductivity of $k_s=1.38$ W$\cdot$ m$^{-1}$ $\cdot$ K$^{-1}$ (the values of these parameters are available in the code~\cite{GWINC} used for thermal noise budget simulations). In the simulations, we considered an incident power $P_{inc} = 750$~kW {\cite{CExplorer,VirgoPRJ}}, a substrate transmissivity $\tau_c=5.6$~ppm, and a total coating absorbance $\alpha_c=1.0$~ppm. Based on the layer-specific absorption shown in Figure~\ref{fig:power}, these parameters result in a volumetric heat source of $A_0=2.985 \times 10^{7}$~W $\cdot$ m$^{-3}$. It is worth noting that such high $P_{inc}$ values are expected in next-generation gravitational wave interferometers, where intracavity laser power will be significantly higher than in current instruments. Moreover, the chosen coating absorbance is higher than that of standard designs. This choice is motivated by studies showing that, when implemented within an optimized multi-material structure, a higher absorbance can lead to a reduction in coating thermal noise~\cite{VP1,VP2,VP3}. The results of these simulations are shown in Figures~\ref{fig:contour}--\ref{fig:increase2}.

Figure~\ref{fig:contour} shows the distribution of temperature increase, with respect to room temperature, on the radial section of the cylinder. A temperature peak can be seen along the TTM axis due to the Gaussian profile of the laser beam and the distance from the lateral surface where thermal dissipation by radiative heat transfer occurs. For a given value of the radial coordinate, the highest temperature level occurs close to the cavity/coating interface, where the laser beam is incident.

This is better highlighted in Figures~\ref{fig:increase1} and~\ref{fig:closeup}, showing the temperature profile on the $r=0$ axis and a close-up of the $0<z<2d$ area (i.e., around the coating). The temperature peak is indeed within the coating, due to the cooling effect of the radiative boundary condition at the surface $z=0$, but the temperature variation within this minuscule thickness is negligible (as displayed in Figure~\ref{fig:closeup}). It is worth noting that the temperature peak occurs in the substrate for the low-absorption binary coatings currently used in detectors~\cite{Rocchi2012}.
 
This behaviour suggests an even stronger homogenisation model, achieved by incorporating the thin coating layer into the boundary condition on the $z=0$ cylinder face, as discussed in the next section.

Finally, Figure~\ref{fig:increase2} gives us the temperature increments on the upper (red) and lower (green) cylinder faces.

Let us conclude with a general comment: The temperature rise is fairly constant in the central TTM zone, near to the cylinder axis (variation $\sim 0.8$ K). The observed temperature rise of approximately $3$ K significantly exceeds the 
sub-$1$K increases typical of scenarios with lower cavity powers and reduced coating absorption. This poses a more challenging thermal management issue, which must be addressed in the system design.

\section{Reduced Thermal Model and Comparative Analysis}
\label{sec:reduced_model_comparison} 

The objective of this section is to simplify the thermal analysis by representing the entire effect of the thin coating---including its internal heat generation and surface convection---as a single, effective boundary condition applied to the main material (the homogeneous cylindrical substrate) at the interface $z=d$. This approach avoids the need to explicitly solve for the temperature distribution within the coating itself, reducing significantly the computational complexity of the problem.

A simplified model can be employed by imposing the following boundary condition at $z=d$:
\begin{equation}
    -k_m \parder{T_m(r,z)}{z}\Big|_{z=d} = h_{eff} (T_{\infty} - T_m(r,d)) + Q_{eff}(r)
\label{eq:reduced}
\end{equation}
where $h_{eff}$ is the effective heat transfer coefficient, $k_m$ is the thermal conductivity of the main material (substrate), and $T_m(r,z)$ is its temperature. The left-hand side of Equation (\ref{eq:reduced}) represents the heat flux entering the homogeneous cylinder from the coating. This equation defines a Robin-type boundary condition at the $z=d$ interface, governing the heat flux into the substrate.

The reduced model (\ref{eq:reduced}) proposed in~\cite{HV1990ThermalAberrations} was employed to develop analytical solutions for TTM heating.

Simplified thermal models for laminated structures are widely used in the analysis of thermal stresses in optics and electronics~\cite{ThermCompl}. A comprehensive review can be found in~\cite{revlayered}. For the specific case of interferometric cavity optics, the seminal work by Hello and Vinet~\cite{HV1990ThermalAberrations} provides the theoretical basis for Equation~\eqref{eq:reduced}.

Due to the very low value of the extinction coefficient of the material, in this model the heat generation in the substrate is neglected. Here, we show the results obtained by using this model, validating its predictions through a numerical comparison.

This effective boundary condition has a specific physical interpretation. The heat flux entering the substrate (left-hand side) is composed of two primary contributions:
\begin{enumerate}
    \item The total heat generated per unit area within the coating, represented by the surface source term $Q_{eff}(r)$. The expression $Q_{\text{eff}}(r)$ is derived from the integration of the volumetric power dissipation (see Equation (\ref{eq:heat_generation})), associated with the incident electromagnetic field, over the thickness $d$ of the coating. We get
    \[
    Q_{eff}(r) = A_0 e^{-2 r^2/w^2} \left( \frac{1 - e^{-\alpha_0 d}}{\alpha_0} \right),
    \]
    where $A_0$ is related to the peak intensity of heat generation, $w$ is the characteristic width of the heat source, and $\alpha_0$ is the absorption coefficient within the coating of thickness $d$.
    \item The heat exchange occurring at the outer surface of the coating ($z=0$). The term $h_{eff} (T_m(r,d) - T_\infty)$ accounts for this heat exchange, influencing the net flux at $z=d$. We assume that the effective heat transfer coefficient $h_{eff}$ is approximately equal to $h_0$, the convection coefficient at the $z=0$ surface.
\end{enumerate}

The numerical solutions for both thermal models were obtained using the finite element method, implemented within the FEniCS computing platform~\cite{fenics_paper}. We now compare the two thermal models: \textbf{Model 1} (reference model) described in Sections 2, 3, and 4 that incorporates a volumetric heat source distributed within a very thin coating layer, extending from $z=0$ to $z=d$, where $d = 8415$\,nm ($\sim$8\,$\upmu$m), and \textbf{Model 2} (simplified model) that employs the reduced boundary condition described by Equation (\ref{eq:reduced}) at the interface $z=d$. The computation time for the simplified model is approximately 3\% of that required by the reference model.

\begin{figure}[H]
	\centering
	\includegraphics[width=0.8\textwidth]{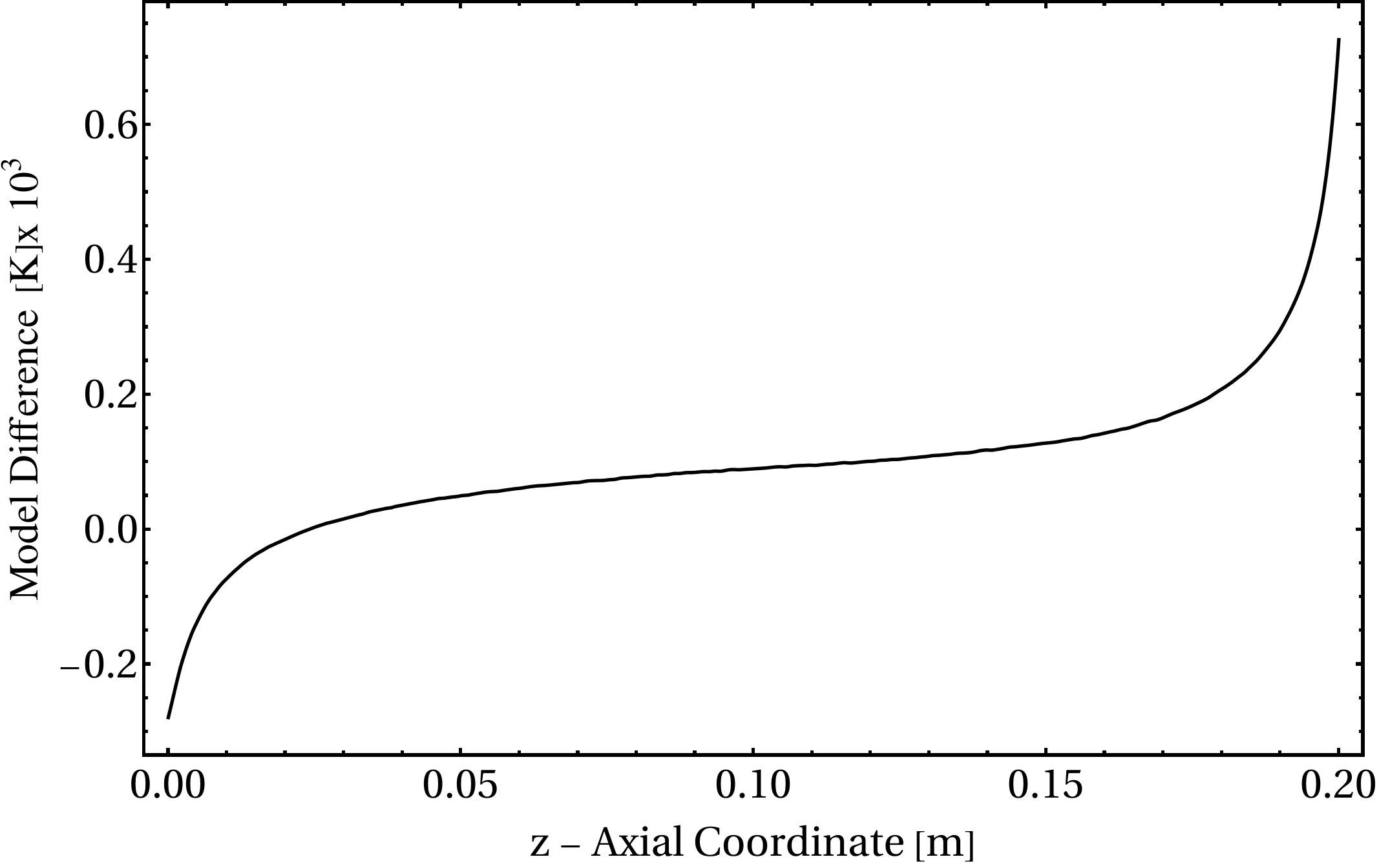}
	\caption{The axial temperature difference ($(\Delta T_{\text{Model 1}} - \Delta T_{\text{Model 2}})$) between the reference volumetric source model (Model 1) and the simplified surface boundary condition model (Model 2) along the cylinder axis.}
	\label{fig:error2}
\end{figure}

Figure~\ref{fig:error2} presents the calculated axial temperature difference between these two models for a substrate of thickness $L=0.2$ m. The y-axis, labeled Model Difference, quantifies $(\Delta T_{\text{Model 1}} - \Delta T_{\text{Model 2}})$. A positive value indicates that Model 1 predicts a higher temperature, while a negative value signifies that Model 2 predicts a higher temperature. The x-axis represents the axial coordinate $z$.

Although the differences between the two models came out to be very small, some insight into the origin of the differences can be gained. Several key observations can be drawn from Figure~\ref{fig:error2}. Near the origin the temperature difference is negative, reaching approximately $-0.3$ mK. This indicates that the simplified model (Model 2) predicts a slightly higher temperature in the immediate vicinity of the coating. Such an outcome is plausible, as concentrating the entire heat input onto the $z=d$ surface in Model 2 can lead to a more intense local temperature peak compared to distributing the heat, even within a very thin volume, as in Model 1.

The curve crosses the zero-difference line, where both models yield identical temperature predictions, at an axial position between $z=0.015$\,m and $z=0.020$\,m. However, up to a distance of $0.15$ m, the difference keeps a small value around $+0.1$ mK. In this region, the reference model (Model 1) predicts slightly higher temperatures. This might be attributed to the concentrated heat application at the $z=d$ surface in Model 2, potentially leading to a slightly more effective initial heat dissipation pathway and thus marginally lower temperatures deeper within the substrate compared to Model 1.

Towards the bottom of the substrate ($z$ approaching $L=0.2$\,m), the positive difference increases, reaching approximately $+0.75$ mK. This deviation indicates that the simplified model (Model 2) underestimates the temperature at this extremity. 

\begin{figure}[H]
	\centering
	\includegraphics[width=0.8\textwidth]{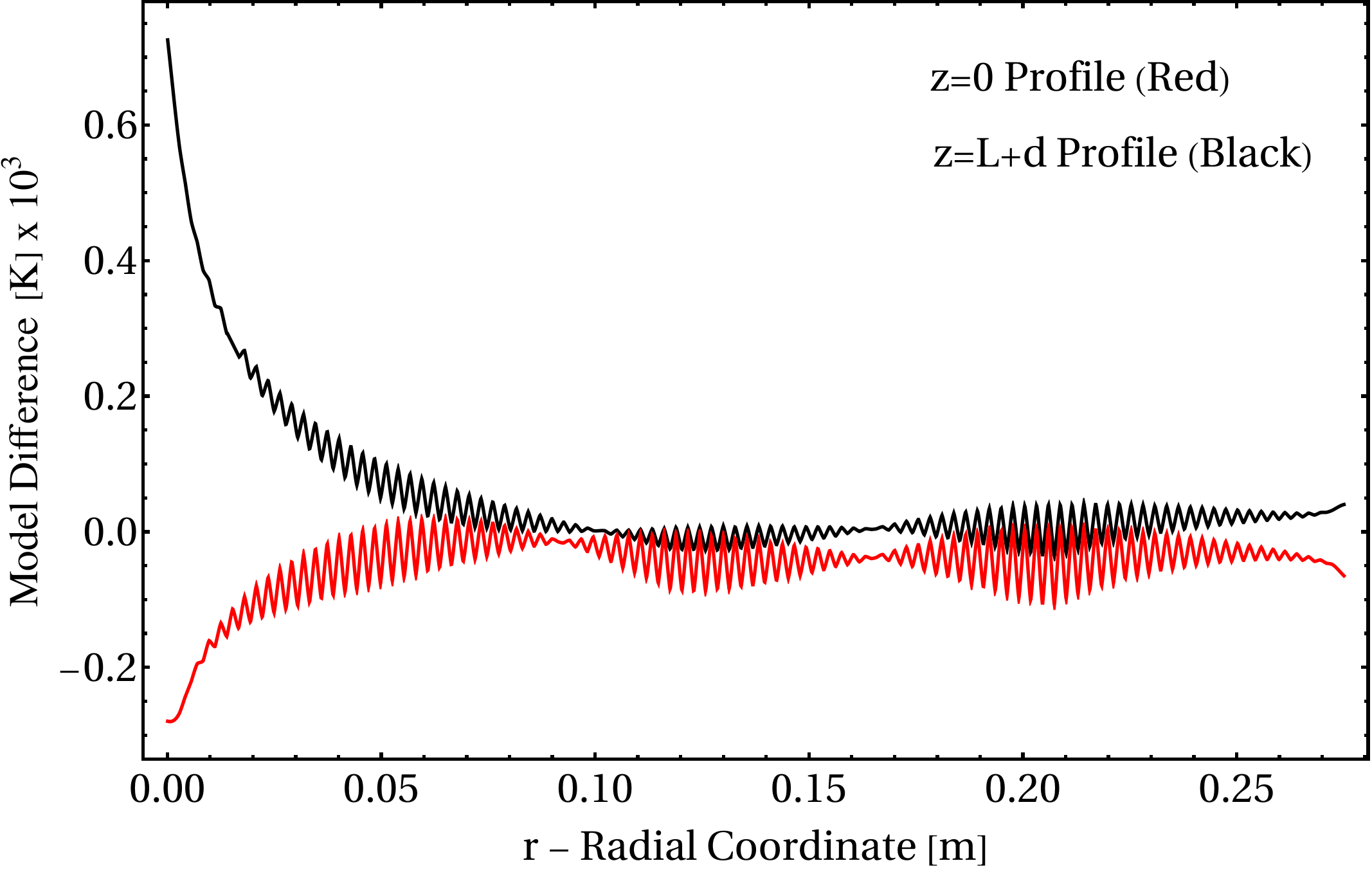}
	\caption{The radial temperature difference ($(\Delta T_{\text{Model 1}} - \Delta T_{\text{Model 2}}) \times 10^3$) between the reference volumetric source model (Model 1) and the simplified surface boundary condition model (Model 2). Profiles are shown for the front surface ($z=d$, red curve) and the back surface ($z=L+d$, black curve).}
	\label{fig:error1}
\end{figure}

Figure~\ref{fig:error1} extends the comparative analysis by displaying the temperature difference between the reference volumetric source model (Model 1) and the simplified surface boundary condition model (Model 2) as a function of the radial coordinate $r$. The x-axis is the radial coordinate, extending to the cylinder radius $R_{cyl}= 0.275$\,m. Two distinct profiles are presented, reflecting the temperature differences on the cylinder's end surfaces:
\begin{itemize}
    \item \textbf{$z = d$ Profile (Red Curve):} This curve shows the radial temperature difference on the front surface of the cylinder ($z=d$), which is immediately adjacent to the very thin coating where heat is primarily introduced or generated in Model 1.
    \item \textbf{$z = L + d$ Profile (Black Curve):} This curve illustrates the radial temperature difference on the back surface of the cylinder, located far from the initial heat source.
\end{itemize}

A conspicuous characteristic of both profiles in Figure~\ref{fig:error1} is the presence of high-frequency oscillations. These are likely numerical artifacts, potentially arising from the mesh resolution at the surfaces or the specific method of data extraction (e.g., point-wise evaluation that might interact with element boundaries). While these oscillations make precise interpretation challenging, the underlying general trends in the model differences can still be discerned.

On the front surface, the temperature difference (Model 1--Model 2) depicted by the red curve starts significantly negative at the centerline ($r=0$), with a value around $-0.3$ mK (i.e., $-0.0003$\,K). This indicates that at the radial center of the front face, the simplified model (Model 2, with its surface flux at $z=d$) predicts a \textit{higher} temperature than the reference volumetric model (Model 1). This outcome is consistent with the axial profile near $z=d$ (Figure~\ref{fig:error2}), where concentrating the heat input as a surface flux at $z=d$ led to a more intense local temperature peak. As the radial coordinate $r$ increases, this negative difference decreases in magnitude (becomes less negative), eventually crossing the zero-difference line around $r \approx 0.07 - 0.08$\,m. 
 
Conversely, on the back surface ($z=L$), the temperature difference (Model 1--Model 2) shown by the black curve begins with a strongly positive value at the centerline ($r=0$), approximately $+0.7$ mK. This signifies that at the radial center of the far end of the cylinder, the reference volumetric model (Model 1) predicts a \textit{higher} temperature than the simplified surface flux model (Model 2). This observation aligns perfectly with the findings from the axial profile (Figure~\ref{fig:error2}) at $z=L$, where Model 2 underestimated the temperature. The volumetric heat distribution in Model 1, even within a very thin layer, appears to result in a heat propagation pattern that maintains higher temperatures along the centerline at greater axial distances. As $r$ increases along this back surface, the positive difference diminishes, crossing the zero-difference line around $r \approx 0.07$\,m. For $r > 0.07$\,m, the difference oscillates around zero, i.e., the magnitude of this difference is small and falls largely within the inaccuracies of the other physical parameters.

The radial profiles provide further insight into the spatial distribution of errors introduced by the simplification:
\begin{itemize}
    \item \textbf{Front Surface ($z=d$):} The simplified model tends to overestimate the temperature at the radial center, likely due to the concentrated nature of the applied surface flux. This overestimation diminishes radially.
    \item \textbf{Back Surface ($z=L+d$):} The simplified model underestimates the temperature at the radial center, confirming the axial analysis. This suggests that the reduced model does not fully capture the axial heat transport along the centerline when the source is volumetric. Moving radially outwards, the discrepancy becomes less pronounced and can even invert.
\end{itemize}

Despite these tiny spatially varying discrepancies, it is crucial to emphasize that the observed temperature differences between the two models are generally on the order of millikelvin. This level of inaccuracy introduced by the simplified model is often significantly overshadowed, if not entirely surpassed, by the inherent experimental uncertainties associated with determining the system's thermo-physical parameters. Such parameters include, but are not limited to, the thermal conductivity of the materials, the emissivity of the surfaces, and the convective and radiative heat transfer coefficients.

\section{Conclusions}
\label{sec:conclusions} 

In this work, we have developed and presented a comprehensive thermal analysis of TTMs featuring a highly reflective ternary coating. Given their potential for next-generation interferometers, these innovative coatings are of significant interest. To the best of our knowledge, the work presented herein constitutes the first detailed investigation of the thermal behaviour of this coating technology. In this framework, we consider operative conditions where TTMs are exposed to high-power laser heating conditions that are expected in next-generation interferometers such as Advanced Virgo and LIGO. Indeed, future instruments will utilize intracavity beam powers significantly higher than those in current detectors.

A primary contribution of this study is the creation and validation of two distinct numerical codes designed to simulate this complex thermal environment. These tools have enabled a detailed investigation into the thermal behavior of TTMs, particularly in scenarios involving higher coating absorption ($\alpha_c = 1$ ppm) that can be potentially considered in future coating designs.

Our simulations, specifically examining a case with a coating absorbance of $1.0$ ppm, have revealed a significant shift in the temperature distribution compared to scenarios with lower absorption (e.g., $0.5$ ppm), which are characteristic of the state-of-the-art coatings in current-generation gravitational wave interferometers~\cite{Virgo_Advanced, LIGO_Advanced}. A key finding is that with increased coating absorption, the peak temperature localizes distinctly near the coating region ($z \approx 0$) rather than in the bulk center of the test mass. This fact highlights the critical role of coating absorption characteristics in dictating the thermal profile and underscores the necessity of accurate modeling of this surface layer, especially as interferometer designs push towards higher power and potentially different coating materials or absorption levels. The resulting temperature increase, on the order of $3$~K under the considered high-power and higher-absorption conditions, would result in an axial TTM length change on the order of $300$~nm, a value consistent with the thermal expansion coefficient of bulk silica $0.5 \times 10^{-6}$\, K$^{-1}$, presenting a more significant thermal management challenge that must be carefully addressed in future thermal control system designs.
 
Furthermore, this study has rigorously compared the results from a detailed model, which explicitly resolves the volumetric heat generation within the thin coating, against a simplified reduced model. The reduced model effectively translates the thermal impact of the coating into an equivalent boundary condition applied to the substrate at the $z=d$ interface. Our comparative analysis demonstrates that while the detailed model provides a more physically complete representation, the differences in temperature predictions between the two approaches are generally on the order of a few millikelvin. This level of discrepancy is often well within the range of, or even surpassed by, the inherent experimental uncertainties in determining crucial thermo-physical parameters such as material thermal conductivities, surface emissivities, and heat transfer coefficients.

Consequently, a significant outcome of this work is the demonstration that an overly detailed and computationally intensive modeling of the coating's internal structure and the precise electromagnetic field distribution (and thus the heat source) within it may not always be necessary for capturing the dominant thermal effects in the TTM. Simplified models, which represent the coating's effect as an effective surface heat source and thermal resistance, can provide sufficiently accurate predictions for many practical purposes, particularly when the total power deposited per unit area in the coating is well-characterized. This finding suggests that for overall thermal management and first-order thermo-elastic distortion analysis, focusing on the net power absorbed by the coating, rather than its intricate internal distribution, can be a computationally efficient and pragmatically sound approach. This allows for a redirection of computational effort towards other critical aspects of interferometer modeling or towards refining the accuracy of the input thermo-physical parameters, which often pose a greater limitation.
  
Ultimately, the findings of this work are expected to inform the design of the detector’s thermal control system. The simplified model could be useful for rapid parameter studies, aimed at finding the optimal working point of this system.

Future work could extend this analysis to transient thermal effects, explore the impact of different coating designs and material properties more broadly, and integrate these thermal models with opto-mechanical simulations to fully assess the implications for interferometer sensitivity.


\section*{Acknowledgments}
The authors wish to thank Innocenzo M. Pinto and Riccardo De Salvo for the fruitful discussions that have enriched this work.
The authors are grateful to the people of the VIRGO CR\&D and VIRGO-ET projects for suggestions and discussion.
This work has been supported in part by Istituto Nazionale di Fisica Nucleare (INFN).

\appendix
\section*{Appendix: Gaussian Beam Waist Change in the Near-Field Region}
\label{app:near_field_expansion}

We aim to find a first-order approximation for the increase in beam radius, $\Delta w = w(L_1+L) - w(L_1)$, under the conditions that the additional displacement $L$ is much smaller than the initial distance $L_1$ from the waist ($L \ll L_1$), and that the initial distance $L_1$ is much smaller than the Rayleigh range $z_R$ ($L_1 \ll z_R$). The latter condition implies we are in the near-field region of the Gaussian beam, or at least not far from the beam waist.

The first-order Taylor expansion for $\Delta w$ around $L_1$ is given by
\begin{equation}
    \Delta w \approx L \cdot \frac{dw}{dz}\bigg|_{z=L_1} = L \cdot w'(L_1)
    \label{eq:taylor_delta_w_app_nf}
\end{equation}
where $w'(z)$ is the derivative of the beam radius $w(z) = w_0 \sqrt{1 + (z/z_R)^2}$ with respect to $z$.
The derivative is
\begin{equation}
    w'(z) = \frac{w_0 z}{z_R^2 \sqrt{1 + (z/z_R)^2}}.
    \label{eq:w_prime_z_app_nf}
\end{equation}
So, at $z=L_1$,
\begin{equation}
    w'(L_1) = \frac{w_0 L_1}{z_R^2 \sqrt{1 + (L_1/z_R)^2}}.
    \label{eq:w_prime_L1_app_nf}
\end{equation}
Now, we apply the near-field condition $L_1 \ll z_R$. This implies $(L_1/z_R)^2 \ll 1$.
Therefore, the term in the square root can be approximated as
\begin{equation}
    1 + \left(\frac{L_1}{z_R}\right)^2 \approx 1.
\end{equation}
So, its square root becomes
\begin{equation}
    \sqrt{1 + \left(\frac{L_1}{z_R}\right)^2} \approx \sqrt{1} = 1.
    \label{eq:sqrt_approx_app_nf}
\end{equation}
Substituting this approximation \eqref{eq:sqrt_approx_app_nf} into the expression for $w'(L_1)$ \eqref{eq:w_prime_L1_app_nf}, we get
\begin{align}
    w'(L_1) &\approx \frac{w_0 L_1}{z_R^2 \cdot 1} \nonumber \\
    w'(L_1) &\approx \frac{w_0 L_1}{z_R^2}.
    \label{eq:w_prime_L1_near_field_app_nf}
\end{align}

Finally, substituting this simplified $w'(L_1)$ from Equation~\eqref{eq:w_prime_L1_near_field_app_nf} into the Taylor expansion for $\Delta w$ from Equation~\eqref{eq:taylor_delta_w_app_nf}, we get the first-order approximation for $\Delta w$ when $L_1 \ll z_R$ (and $L \ll L_1$):
\begin{equation}
    \Delta w \approx L \cdot \frac{w_0 L_1}{z_R^2}.
    \label{eq:delta_w_near_field_app_nf}
\end{equation}

This result shows that in the near-field region or close to the waist (where $L_1 \ll z_R$), the change in beam radius $\Delta w$ for a small additional propagation distance $L$ is proportional to $L$, $L_1$, and $w_0$, but inversely proportional to $z_R^2$. The factor $L_1/z_R$ is small by our initial assumption ($L_1 \ll z_R$). The term $w_0/z_R$ is related to the far-field divergence angle. Thus, the overall rate of change of the waist $w'(L_1) \approx (w_0/z_R) \cdot (L_1/z_R)$ is significantly suppressed compared to the far-field divergence. This indicates that the beam radius changes very slowly for small displacements $L$ when the starting point $L_1$ is close to the beam waist or well within the Rayleigh range. This is consistent with the behavior of a Gaussian beam, which is most collimated (i.e., its radius changes slowest) around its waist.

For gravitational wave interferometers like Virgo/LIGO, the calculated change in beam radius is approximately $(10^{-13}-10^{-15})$ m, extremely small. 


\end{document}